\documentclass[english,12pt,aps,prd,a4paper,preprintnumbers,floatfix,nofootinbib,showpacs,superscriptaddress]{revtex4-1}
\usepackage{amsmath}
\usepackage{amstext}
\usepackage{amssymb}
\usepackage{graphicx}
\usepackage{hyperref}
\usepackage{rotating}
\usepackage{gensymb}
\usepackage[normalem]{ulem} 
\usepackage[utf8]{inputenc}
\usepackage{color} 
\usepackage{multirow} 
\usepackage[T1]{fontenc} 
\usepackage{slashed}
\usepackage{caption}
 \usepackage[sort&compress]{natbib} 
\usepackage[english]{babel}
\usepackage{amsmath}
\usepackage{graphicx}
\usepackage{color}
\usepackage{mathrsfs}   
\usepackage{amssymb}
\usepackage{hyperref}
\usepackage{dcolumn}
\usepackage{bm}
\usepackage{graphicx}
\usepackage{mathrsfs}   
\usepackage{slashed}     
\usepackage{url} 
\usepackage{graphicx}
\usepackage{float}
\usepackage{caption}
\usepackage{diagbox}
\usepackage{makecell}
\usepackage{pifont}

\definecolor{greenLinks}{rgb}{0, 0.6, 0} 
\definecolor{blueLinks}{rgb}{0, 0, 0.6}
\definecolor{redLinks}{rgb}{0.6, 0, 0}
\definecolor{tempText}{rgb}{0.55, 0.10,0.67}
\definecolor{eprintLinks}{rgb}{0.4, 0.4, 0.4}
\definecolor{journalLinks}{rgb}{0.6, 0, 0}

\newcommand{\MYhref}[3][redLinks]{\href{#2}{\color{#1}{#3}}}%

\usepackage{float}

\newcommand{\cmark}{\ding{51}}%
\newcommand{\xmark}{\ding{55}}%

\def\gsim{\raise0.3ex\hbox{$\;>$\kern-0.75em\raise-1.1ex\hbox{$\sim\;$}}}
\def\lsim{\raise0.3ex\hbox{$\;<$\kern-0.75em\raise-1.1ex\hbox{$\sim\;$}}}

\def\21{$\mathrm{SU(2)_L \otimes U(1)_Y}$ }

\newcommand{\eVq}  {\text{eV}^2}
 \usepackage[sort&compress]{natbib}
\newcommand{\AddrAHEP}{%
  AHEP Group, Institut de F\'{i}sica Corpuscular --
  C.S.I.C./Universitat de Val\`{e}ncia, Parc Cientific de Paterna.\\
 C/ Catedratico Jos\'e Beltr\'an, 2 E-46980 Paterna (Val\`{e}ncia) - SPAIN}

\begin{document}

\title{ Zooming in on neutrino oscillations with DUNE} 

\author{Rahul Srivastava}\email{rahulsri@ific.uv.es}
\author{Christoph A. Ternes}\email{chternes@ific.uv.es}
\author{Mariam T{\'o}rtola}\email{mariam@ific.uv.es}  
\author{Jos\'e W.~F.~Valle} \email{valle@ific.uv.es} 
\affiliation{\AddrAHEP}

\begin{abstract}
 \vspace{1cm} We examine the capabilities of the DUNE
  experiment as a probe of the neutrino mixing paradigm.
  Taking the current status of neutrino oscillations and the design
  specifications of DUNE, we determine the experiment's potential to
  probe the structure of neutrino mixing and CP violation.
  We focus on the poorly determined parameters $\theta_{23}$ and
  $\delta_{CP}$ and consider both two and seven years of run.
  We take various benchmarks as our true values, such as the current
  preferred values of $\theta_{23}$ and $\delta_{CP}$, as well as
  several theory-motivated choices.
  We determine quantitatively DUNE's potential to perform a precision
  measurement of $\theta_{23}$, as well as to test the CP violation
  hypothesis in a model-independent way.
  We find that, after running for seven years, DUNE will make a
  substantial step in the precise determination of these parameters,
  bringing to quantitative test the predictions of various theories of
  neutrino mixing.
  
 \end{abstract}

\maketitle

\section{Introduction}
\label{sec:intro}

Ever since the confirmation of the experimental discovery of neutrino
oscillations~\cite{Kajita:2016cak,McDonald:2016ixn} there has been
a flood of studies, both experimental and theoretical.
Indeed, many experimental studies have been conducted, and it is fair
to say that oscillation experiments have probed many of the key
features of the oscillation picture, summarized in the global fit
results given in Ref.~\cite{deSalas:2017kay}.
Though we still lack precise information on leptonic CP violation, the
neutrino mass ordering and the octant of the atmospheric mixing angle
$\theta_{23}$, we have pretty good information on the remaining
oscillation parameters.

On theoretical side, there have been many attempts to understand the
physics associated to the origin of neutrino mass as well as to shed
light on the pattern of neutrino mixing. 
In particular, the approach of flavor symmetries to explain the
observed neutrino oscillation data has been widely
used~\cite{Morisi:2012fg,King:2014nza}.
For example, the precise measurement of the non-zero reactor angle
$\theta_{13}$ has ruled out many proposals for neutrino mixing
pattern, such as the celebrated tri-bimaximal (TBM) mixing ansatz,
characterized by the Harrison-Perkins-Scott lepton mixing
matrix~\cite{harrison:2002er}.
Likewise, it has ruled out well--motivated theories of neutrino mass,
such as the minimal Babu-Ma-Valle (BMV) model~\cite{Babu:2002dz},
subsequently revamped into~\cite{Morisi:2013qna,Chatterjee:2017ilf}.

The search for neutrino oscillations at the upcoming long baseline
experiments, such as the Deep Underground Neutrino Experiment (DUNE),
will play a key role in the agenda of neutrino physics experimentation
over the coming decades~\cite{Acciarri:2016ooe,Acciarri:2015uup}.
It will be able to substantially improve our current measurement of the
$\theta_{23}$ angle and can potentially provide a precise measurement
of $\delta_{CP}$ the leptonic CP phase.  Thus it can test various
leptonic mixing models and can provide an enhanced understanding of
the physics behind it.
Our paper is structured as follows. In section~\ref{sec:benchmark} we
give the description and motivation for the benchmark points used in
our paper. These include both specific points in the
$\delta_{CP}-\theta_{23}$ plane, 
  subsections~\ref{sec:exp-benchmark-points} and
  \ref{sec:theory-benchmark-points}, as well as lines in that plane,
in subsection~\ref{sec:benchmark-lines}.
In section~\ref{sec:simulation} we describe the details of our
simulation of the DUNE experiment. 
Our results are presented in section~\ref{sec:results}, where we
provide a detailed explanation for all the analyzed cases, see
subsections~\ref{sec:point-benchmarks} and~\ref{sec:line-benchmarks}.
  Finally, in Table~\ref{tab:summary}, given in
  section~\ref{sec:summary-discussion}, we give an ``executive''
  summary of our results.\\[-.2cm]

\section{Benchmarks}
\label{sec:benchmark}

Motivated by the potential of the DUNE experiment to probe CP
violation and substantially improve the precision in the determination
of neutrino oscillation parameters,  
we examine some of the well--motivated and popular
proposals that can be tested at DUNE.
Our benchmarks are listed in Tabs.~\ref{tab:benchmark-points-exp},
\ref{tab:benchmark-points-theory} and \ref{tab:benchmark-lines} and
are divided into three broad categories.  Our first category
(see Sec.~\ref{sec:exp-benchmark-points}) consists of the
experimentally motivated benchmarks i.e. the current best fit points
and local minima obtained from global fits of neutrino oscillation
data~\cite{deSalas:2017kay}.
In Sec.~\ref{sec:theory-benchmark-points} we look at theoretical
predictions for $\theta_{23}$ and $\delta_{CP}$ that are often used in
the literature. These benchmark points are motivated by some of
  the popular theoretical scenarios for the pattern of neutrino
  mixing~\cite{harrison:2002er,Babu:2002dz,harrison:2002kp,Altarelli:2005yx,Ma:2005qf,deMedeirosVarzielas:2005qg,harrison:2002kp,grimus:2003yn,Datta:2003qg,Rodejohann:2008ir,Ma:2016nkf,
    Ma:2017trv, Ma:2017moj}.
  In Sec.~\ref{sec:benchmark-lines} we take a more general approach
  and examine the potential of DUNE as a probe of the maximality of
  the $\theta_{23}$ angle, irrespective of the $\delta_{CP}$ value,
  and of maximal (or null) CP violation, irrespective of the
  $\theta_{23}$ value. These benchmarks provide useful guiding posts
  once the DUNE experiment will start collecting data.
We now give a brief description and motivation for the benchmark
points used in this paper. We also indicate the figures summarizing
the results of our simulation. Their detailed explanation is given
in Sec.~\ref{sec:results}.\\[-.2cm]


\subsection{Experimental Benchmark points}
\label{sec:exp-benchmark-points}


Here we discuss a number of benchmark points which are directly
motivated by the current experimental data on the leptonic
mixing~\cite{deSalas:2017kay}.

\subsubsection{Global minimum for normal mass ordering}

The global fit of current neutrino oscillation data indicates
that, if neutrinos have normal mass ordering, then the best fit after
combining all of the data corresponds to
$\sin^2 \theta_{23} = 0.430$ and $\delta_{CP} = 1.40 \pi$. Motivated by the
current experimental status we examined the possibility of probing the
unknown values of the oscillation parameters
$\theta_{23}$ and $\delta_{CP}$, taking the current best fit point 
value as the true value chosen by nature. The result of our DUNE
simulation for this case is shown in the left panel of
Fig.~\ref{fig:sq23-del-GLOBAL-FIT-NO}.

\subsubsection{Local minimum for normal mass ordering}

In addition to the global best fit point mentioned above, the $\chi^2$ function
has a local minimum in the upper octant of $\theta_{23}$, corresponding
to $\sin^2 \theta_{23} = 0.596$ and $\delta_{CP} = 1.16 \pi$. 
Since the current data are not enough to discard this possibility in a
significant way, we regard it as viable benchmark point and examine
the possibility of probing the unknown oscillation parameters
$\theta_{23}$ and $\delta_{CP}$ taking this point as the true value.
 The result of our DUNE simulation for this case is
shown  in the right panel of Fig.~\ref{fig:sq23-del-GLOBAL-FIT-NO}.

\subsubsection{Global Minimum for inverted mass ordering}

We consider the minima obtained in inverted mass ordering (IO) also as
viable benchmark points. In this case, the global minimum of the fit
lies in the second octant of $\theta_{23}$, with values corresponding
to $\sin^2 \theta_{23} = 0.598$ and $\delta_{CP} = 1.56 \pi$. Since the
mass ordering of neutrinos is still unknown~\cite{Gariazzo:2018pei},
we regard this possibility as another viable choice for the true value
for our simulation. The result of the DUNE simulation corresponding to
this benchmark point
is shown in the left panel of Fig.~\ref{fig:sq23-del-GLOBAL-FIT-IO}. 

\subsubsection{Local minimum for inverted mass ordering}

Also for inverted mass ordering there is a local minimum, but now
located in the first octant of $\theta_{23}$, corresponding to
$\sin^2 \theta_{23} = 0.425$ and $\delta_{CP} = 1.52 \pi$. Again, we have
taken this possibility as the fourth possible ``experimental''
benchmark point, showing our results in the right panel 
of Fig.~\ref{fig:sq23-del-GLOBAL-FIT-IO}. \\[-.2cm]

The values of $\sin^2 \theta_{23}$ and $\delta_{CP}$ associated
  to our experimentally motivated benchmark points are summarized in
  Tab.~\ref{tab:benchmark-points-exp}.
\begin{table}[h!]\centering
  \catcode`?=\active \def?{\hphantom{0}}
   \begin{tabular}{|l|c|c|}
    \hline
    Motivation & $\sin^2\theta_{23}$ & $\delta_{CP}/\pi$
    \\
    \hline
    Global Minimum (NO), Fig.~\ref{fig:sq23-del-GLOBAL-FIT-NO}    & 0.430        &  1.40  \\  
    Local Minimum (NO), Fig.~\ref{fig:sq23-del-GLOBAL-FIT-NO}     & 0.596        &  1.16  \\  
    Global Minimum (IO), Fig.~\ref{fig:sq23-del-GLOBAL-FIT-IO}    & 0.598        &  1.56 \\
    Local Minimum (IO), Fig.~\ref{fig:sq23-del-GLOBAL-FIT-IO}      &  0.425          &  1.52 \\
\hline
     \end{tabular}
          \captionsetup{justification=raggedright}
          \caption{ Experimentally motivated benchmark points~\cite{deSalas:2017kay}.}
     \label{tab:benchmark-points-exp} 
\end{table}


\subsection{Theoretical Benchmark points}
\label{sec:theory-benchmark-points}


\subsubsection{Maximal atmospheric mixing and CP conservation
  with $\delta_{CP} = 0$}
  
Maximal atmospheric mixing is a generic prediction of several leptonic
mixing matrix ansatzes. Here we consider the BMV
  model~\cite{Babu:2002dz}, as well as schemes with the TBM mixing
  pattern~\cite{harrison:2002kp,Altarelli:2005yx, Ma:2005qf,
    deMedeirosVarzielas:2005qg}, and the celebrated $\mu -\tau$
  symmetry~\cite{harrison:2002kp,grimus:2003yn}.
  Maximal $\theta_{23}$ also emerges for the Grimus-Lavoura (GL)
  version of BMV~\cite{grimus:2003yn}, the Golden Ratio
  (GR)~\cite{Datta:2003qg,Rodejohann:2008ir}, as well as co-bimaximal
  mixing (CB) schemes~\cite{Ma:2016nkf, Ma:2017trv, Ma:2017moj}, and
  is often accompanied by the prediction of CP conservation.
  Notice that several of the above scenarios, such as TBM and BMV have
  $\theta_{13}=0$ and are at odds with reactor data from Daya
  Bay~\cite{An:2016ses}, RENO~\cite{Pac:2018scx} and Double
  Chooz~\cite{Abe:2014bwa}.  However they can be generalized so as to
  be consistent with data. For instance, the ``revamped'' BMV model of
  Ref.~\cite{Morisi:2013qna} can be considered on its own right and it
  has been contrasted with oscillation data in a dedicated
  manner~\cite{Chatterjee:2017ilf}.  It is useful, however, to examine
  the simplest ``unrevamped'' TBM and BMV benchmark points.
Having this as motivation we have also analyzed various benchmark
scenarios corresponding to maximal $\theta_{23}$, such as the
theoretical benchmark point
($\sin^2 \theta_{23} = 0.5, \delta_{CP} = 0$). The DUNE simulation corresponding to this possibility 
is shown in the left panel of Fig.~\ref{fig:sq23-del-TBM}. 

\subsubsection{Maximal atmospheric mixing and CP conservation
    with $\delta_{CP} = \pi$}
  
  This is the other benchmark point for the case of maximal
  $\theta_{23}$ and no CP violation. Since the case of CP conservation
  implies either $\delta_{CP} = 0$ or $\delta_{CP} = \pi$, we also
  have taken this as an alternative scenario, and present in the right
  panel of Fig.~\ref{fig:sq23-del-TBM} the result of the DUNE
  simulation for this case.

\subsubsection{Maximal atmospheric mixing and maximal CP violation with $\delta_{CP} = \pi/2$}
  
Some work in the literature predicts maximal $\theta_{23}$ and maximal
CP violation~\cite{Ma:2016nkf}.
Since maximal CP violation implies either $\delta_{CP} = \pi/2$ or
$\delta_{CP} = 3 \pi/2$, we have two options for this benchmark.
Although disfavored, current oscillation data do not exclude maximal
CP violation with $\delta_{CP} = \pi/2$.  The result of our DUNE
simulation obtained for this case is shown in the left panel of
Fig.~\ref{fig:sq23-del-COBI}.

\subsubsection{Maximal atmospheric mixing and maximal CP violation with $\delta_{CP} = 3\pi/2$}
    
The global fit of neutrino oscillation experiments suggests that
leptonic CP violation is maximal, characterized by
$\delta_{CP} \approx 3\pi/2$ as the preferred value.  Motivated by the
experimental hint, we have examined this possibility.  The result of
the DUNE simulation for $\sin^2 \theta_{23} = 0.5$ and
$\delta_{CP} = 3\pi/2$ is shown in the right panel of
Fig.~\ref{fig:sq23-del-COBI}.

\subsubsection{Bi-large mixing with $\delta_{CP} = 0$}

The bi-large mixing ansatz is another interesting and somewhat unique
mixing pattern, which aims to relate the leptonic mixing angles with
the Cabbibo angle of the quark
sector~\cite{Boucenna:2012xb,Ding:2012wh,Roy:2014nua}.  It predicts
$\sin^2 \theta_{23} = 0.45$ with an unpredicted value of
$\delta_{CP}$. For the sake of definiteness, here we have taken the
bi-large predicted value of $\theta_{23}$ angle for the case of no CP
violation. Thus, our benchmark point for this case is
($\sin^2 \theta_{23} = 0.45, \delta_{CP} = 0$). The result of the DUNE
simulation for this case is shown in Fig.~\ref{fig:sq23-del-BILARGE}.\\

The values of $\sin^2 \theta_{23}$ and $\delta_{CP}$ associated
  to our theoretically motivated benchmark points are summarized in
  Tab.~\ref{tab:benchmark-points-theory}.
 \begin{table}[!h]\centering
  \catcode`?=\active \def?{\hphantom{0}}
   \begin{tabular}{|l|c|c|}
    \hline
    Motivation & $\sin^2\theta_{23}$ & $\delta_{CP}/\pi$
    \\
    \hline   
    TBM, BMV, $\mu-\tau$, GR, Fig.~\ref{fig:sq23-del-TBM}   & 0.5        &  0.0 \\
    TBM, BMV, $\mu-\tau$, GR, Fig.~\ref{fig:sq23-del-TBM}   & 0.5        &  1.0 \\
    CB, BMV(GL), Fig.~\ref{fig:sq23-del-COBI}       & 0.5        &  0.5 \\
    CB, BMV(GL), Fig.~\ref{fig:sq23-del-COBI}       & 0.5        &  1.5 \\
    Bi-large, Fig.~\ref{fig:sq23-del-BILARGE}                            & 0.45 &  0.0 \\
    \hline
     \end{tabular}
          \captionsetup{justification=raggedright}
          \caption{ Theory motivated benchmark
            points~\cite{harrison:2002er,Babu:2002dz,Morisi:2012fg,King:2014nza}. }
     \label{tab:benchmark-points-theory} 
\end{table}


\subsection{Benchmark Lines}
\label{sec:benchmark-lines}


After discussing the benchmark points described above, we now give a
brief description of the benchmark lines.
These help us to have an idea of the constraining power of the DUNE
experiment, i.e., how much DUNE can constrain leptonic mixing in a
more model independent way.
In the following we present briefly the benchmark lines to be used in
our simulations.

\subsubsection{Maximal atmospheric mixing}

We first consider the benchmark line corresponding to maximal
atmospheric mixing angle, $\sin^2 \theta_{23} = 0.5$, with no definite
fixed value of $\delta_{CP}$.
We have examined the capabilities of the DUNE experiment to probe this
case by performing such model independent simulation, whose
result is shown in Fig.~\ref{fig:sq23-del-LINE-th23}.

\subsubsection{CP conservation with $\delta_{CP} = 0$}

The possibility of leptonic CP violation is one of the most important
questions that DUNE can address. Motivated by this we have studied
model independent scenarios for leptonic CP violation. 
One of the simplest possibilities is that there is no leptonic CP
violation at all. This means that  $\delta_{CP}$ is either equal
  to 0 or $\pi$~\footnote{If neutrinos are Majorana fermions there
  are, apart from $\delta_{CP}$, two Majorana phases which lead to CP
  violation~\cite{Schechter:1980gr}. However these phases cannot be
  probed by neutrino oscillation experiments.}.
Therefore, as one of our line benchmarks 
  we took $\delta_{CP} = 0$ with $\theta_{23}$ varying within a rather conservative range of:
  $\sin^2\theta_{23}\in[0.35,0.65]$. Values of
  the atmospheric mixing angle outside this range are already excluded
  with high statistical significance~\cite{deSalas:2017kay}.
 The result of this simulation is shown in the left panel of Fig.~\ref{fig:sq23-del-LINE-NOCP}.

\subsubsection{CP conservation with $\delta_{CP} = \pi$}

Apart from $\delta_{CP} = 0$, the other possible value of
$\delta_{CP}$ which leads to no leptonic CP violation is
$\delta_{CP} = \pi$. Thus we took this value for arbitrary
$\theta_{23}$ as another benchmark line in our simulations, leading to
the result shown in the right panel of
Fig.~\ref{fig:sq23-del-LINE-NOCP}.

\subsubsection{Maximal CP violation with $\delta_{CP} = \pi/2$}

The possibility of maximal leptonic CP violation is also quite
intriguing. In order to test it we have taken this as a reference
case for our DUNE simulation.
To start we consider the first possibility, namely,
$\delta_{CP} = \pi/2$ with arbitrary $\theta_{23}$. The
results are shown in the left panel of
Fig.~\ref{fig:sq23-del-LINE-MAXCP}.

\subsubsection{Maximal CP violation with $\delta_{CP} = 3\pi/2$}

Maximal CP Violation can also arise when
$\delta_{CP} = 3\pi/2$ with arbitrary $\theta_{23}$. In fact,
recent global oscillations fits favour $\delta_{CP}$ quite close to
this value~\cite{deSalas:2017kay}.
Although this hint is not yet too robust, DUNE can lead to a
significant improvement. We have taken this as our last benchmark
line, for which our simulations give the results shown in the right
panel of Fig.~\ref{fig:sq23-del-LINE-MAXCP}.\\

The ranges of $\sin^2 \theta_{23}$ and $\delta_{CP}$ for the benchmark lines are summarized in Tab.~\ref{tab:benchmark-lines}.

\begin{table}[h!]\centering
  \catcode`?=\active \def?{\hphantom{0}}
   \begin{tabular}{|l|c|c|}
    \hline 
    Motivation & $\sin^2\theta_{23}$ & $\delta_{CP}/\pi$
    \\
    \hline
    Maximal mixing, Fig.~\ref{fig:sq23-del-LINE-th23}         & 0.5                &  $[0,2]$      \\ 
    CP conservation, Fig.~\ref{fig:sq23-del-LINE-NOCP}        & [0.35,0.65]          & 0.0      \\
    CP conservation, Fig.~\ref{fig:sq23-del-LINE-NOCP}        & [0.35,0.65]          & 1.0      \\  
    Maximal CP Violation, Fig.~\ref{fig:sq23-del-LINE-MAXCP}  & [0.35,0.65]          &  0.5      \\
    Maximal CP Violation, Fig.~\ref{fig:sq23-del-LINE-MAXCP}  & [0.35,0.65]          &  1.5      \\
    \hline
     \end{tabular}
          \captionsetup{justification=raggedright}
          \caption{Probing maximal/null CP violation and maximality of
            $\theta_{23}$.}
     \label{tab:benchmark-lines} 
\end{table}

Having reached this point let us comment that, most theories of
  neutrino mixing do not predict specific values for the neutrino
  oscillation parameters, rather they yield regions in the
  $\sin^2\theta_{23}$-$\delta_{CP}$
  plane~\cite{Morisi:2013qna,Chen:2015jta,Chen:2015siy,CarcamoHernandez:2017owh,
    CentellesChulia:2017koy,Bonilla:2017ekt}.
  For such cases, a more meaningful way to confront the given  predictive
  theoretical model with neutrino oscillation data is to preform a
  dedicated constrained $\chi^2$-fit. Although in its infancy, this
  program has been carried out for a number of theories of lepton
  mixing~\cite{Pasquini:2016kwk,Chatterjee:2017ilf,
    Chatterjee:2017xkb,Srivastava:2017sno}.

\section{Simulation of the DUNE experiment}
\label{sec:simulation}

The Deep Underground Neutrino Experiment (DUNE) is a large-scale
international collaboration aiming to detect neutrinos a mile
underground beneath an abandoned gold mine located in South Dakota, at
about 800 mile (1300 km) distance from their production site at
Fermilab, in Batavia, Illinois.
DUNE is expecting around $1.47\times 10^{21}$ protons on target each
year, due to its 80 GeV beam with 1.07 MW beam power, considerably
more than the present-day experiments
T2K~\cite{Abe:2017bay,Abe:2017uxa} and
NO$\nu$A~\cite{Adamson:2017qqn,Adamson:2017gxd}.

In order to simulate DUNE we use the GLoBES package~\cite{Huber:2004ka,Huber:2007ji} 
together with the auxiliary file presented in Ref.~\cite{Alion:2016uaj}.
Our simulation of DUNE considers a period of 1 as well as 3.5 years running time in
both neutrino and antineutrino mode, taking into account the
disappearance and appearance channels for neutrinos and antineutrinos.
Following Refs.~\cite{Acciarri:2015uup} and~\cite{Alion:2016uaj} we
include several types of background events.  These are due to
misinterpretation of neutrinos as antineutrinos and vice-versa,
contamination of electron neutrinos and antineutrinos in the beam,
misinterpretation of muon as electron neutrinos, as well as the
appearance and misinterpretation of tau neutrinos and neutral current
interactions.  We associate to each of the backgrounds a nuisance
parameter, ranging between 5\% and 20\%, over which we later
marginalize.  In addition, we assign a 2\% error on the signals in
the appearance channels and a 5\% error in the disappearance channels,
as indicated in the studies performed by the DUNE Collaboration in
Ref.~\cite{Acciarri:2015uup}.

In this work we will be mainly interested in the worse determined
oscillation parameters $\sin^2\theta_{23}$ and $\delta_{CP}$,
therefore we simulate the future event rate in DUNE fixing the other
parameters to their best fit values reported in
\cite{deSalas:2017kay}.
In order to determine the DUNE sensitivity to the parameters of
interest, we then marginalize over $\theta_{13}$, $\theta_{12}$,
$\Delta m_{31}^2$ and $\Delta m_{21}^2$ within their 1$\sigma$-ranges,
see Table~\ref{tab:params-margs}.
We generate future DUNE data for several pairs of
$(\theta_{23}^\text{true},\delta_{CP}^\text{true})$ motivated by
experiments and models (see
Tabs~\ref{tab:benchmark-points-exp},\ref{tab:benchmark-points-theory},\ref{tab:benchmark-lines}).
    For each set of reconstructed parameters
    $(\theta_{23},\delta_{CP})$ we then calculate the
    $\chi^2$-function, given as
\begin{equation}
 \chi^2(\theta_{23},\delta_{CP})=
 \min_{\theta_{1j}, \Delta m_{j1}^2,\vec{\alpha}}\sum_\text{channels} 2\sum_n\left[ N_n^{\text{test}}- N_n^{\text{dat}} +
 N_n^{\text{dat}}\log\left(\frac{N_n^{\text{dat}}}{N_n^{\text{test}}}\right)\right] 
 + \sum_i \left(\frac{\alpha_i}{\sigma_i}\right)^2 ,
 \label{main-chi2}
\end{equation}
where $\theta_{1j}$ and $\Delta m_{j1}^2$ ($j=2,3$) denote the four
well-measured oscillation parameters.
Here $N_n^\text{dat}$ corresponds to the simulated event number in the
$n$-th bin obtained with $\theta_{23}^\text{true}$ and
$\delta_{CP}^\text{true}$.
$N_n^{\text{test}}$ is the event number in the $n$-th bin associated
to the parameters $(\theta_{23},\delta_{CP})$ and $\alpha_i$
and $\sigma_i$ are the nuisance parameters and their corresponding
standard deviations, respectively.
Note that $N_n^{\text{test}}$ also depends on $\vec\alpha$, since
these can change the number of signal and background events.
Finally, we sum the $\chi^2$-grid in the $\delta_{CP}-\theta_{23}$ plane from the global fit~\cite{deSalas:2017kay}, to include our current knowledge on those parameters.
\begin{table}[t!]\centering
  \catcode`?=\active \def?{\hphantom{0}}
   \begin{tabular}{|l|c|c|}
    \hline
    parameter & best fit value & relative error
    \\
    \hline
    $\Delta m^2_{21}$& $\phantom{-}7.56\times 10^{-5}\eVq$&2.5\%\\  
    $\Delta m^2_{31}$ (NO)&  $\phantom{-}2.55\times 10^{-3}\eVq$&1.6\%\\
    $\Delta m^2_{31}$ (IO)&  $-2.47\times 10^{-3}\eVq$&1.6\%\\
    $\sin^2\theta_{13}$ & 0.02155&3.9\%\\
    $\sin^2\theta_{12}$ & 0.321&5.5\%\\
    \hline
     \end{tabular}
          \captionsetup{justification=raggedright}
          \caption{ Best fit values and 1$\sigma$ relative
            uncertainties for the better determined neutrino
            oscillation parameters from \cite{deSalas:2017kay}.}
     \label{tab:params-margs} 
\end{table}

\section{Results}
\label{sec:results}

In this section we present our main results for the chosen benchmarks
described above. As explained in Sec.~\ref{sec:simulation}, the results
presented in the figures are obtained by taking into account our
current knowledge of $\theta_{23}$ and $\delta_{CP}$ by adding the
corresponding $\chi^2$-grid from Ref.~\cite{deSalas:2017kay}. 
We have performed simulations of DUNE for two years and for seven
years of running time, divided equally between neutrino and
antineutrino modes in both cases.
The assumed true value in each plot is denoted by a star for quick visual
reference.
We plot the expected regions for two years running time in blue and
for seven years of running time in red.
Moreover, for both cases we show our results for 3$\sigma$ (dashed
lines) as well as 5$\sigma$ (solid lines) confidence levels. For
definiteness, we have assumed normal neutrino mass ordering for all of
our theory-motivated benchmarks~\footnote{The results for inverted
  mass ordering can be obtained in a straightforward way. They are
  very similar and the conclusions do not differ significantly.}.

\subsection{Benchmark points}
\label{sec:point-benchmarks}

\begin{figure}[!t]
 \centering
      \includegraphics[width=0.7\textwidth]{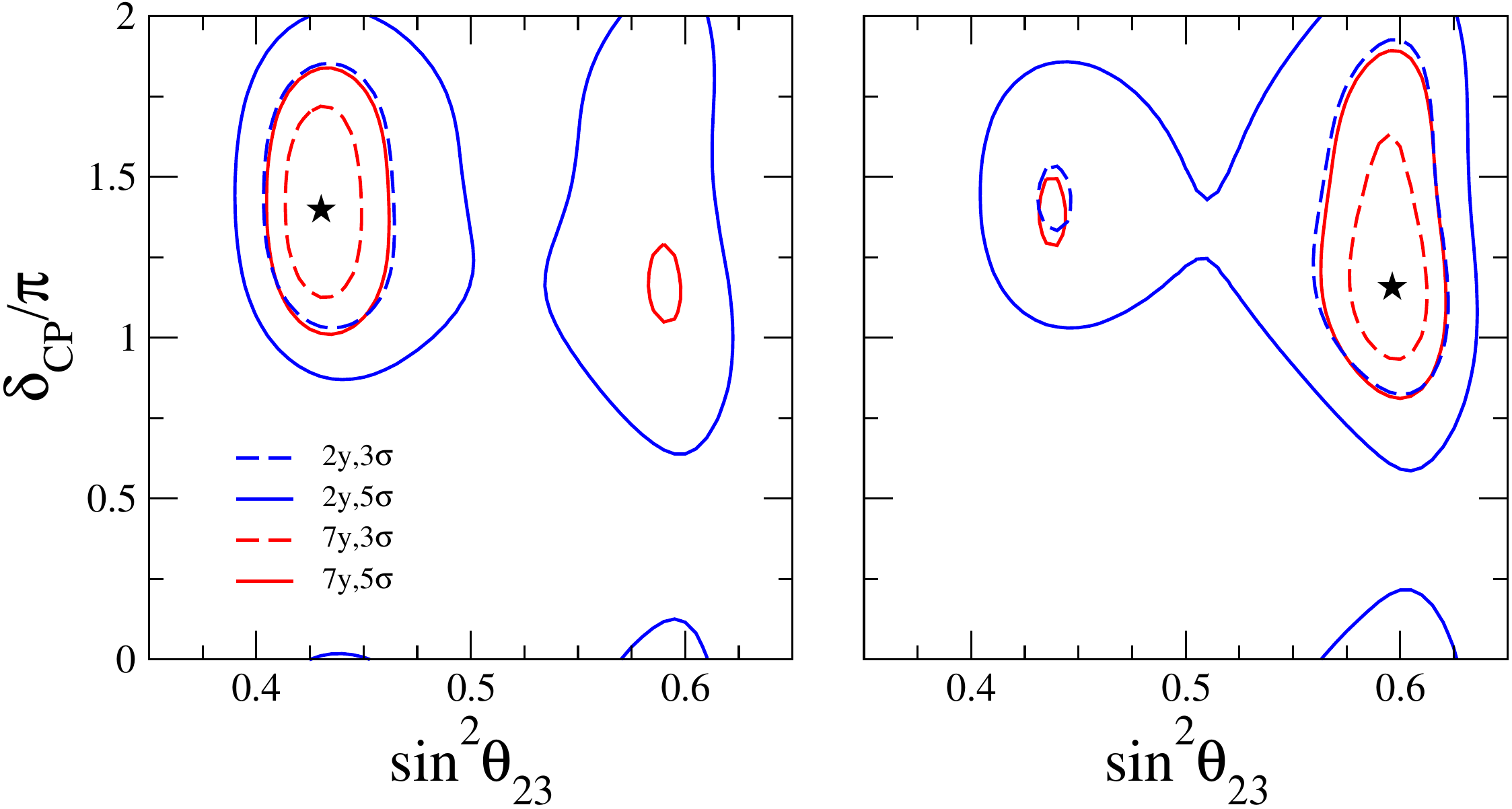}
      \captionsetup{justification=raggedright}
      \caption{ DUNE sensitivity projections for normal mass
        ordering of neutrinos, after 2 years (blue) and 7 years (red) run, taking
        the global best fit point (left panel) and the local minimum
        (right panel) of Ref.~\cite{deSalas:2017kay} as benchmark
        points (shown in star). The dashed lines represent $3\sigma$
        and solid lines are for $5\sigma$ C.L. sensitivities.}
	\label{fig:sq23-del-GLOBAL-FIT-NO}
\end{figure}

We start by taking the experimentally motivated benchmark points
presented in Tab.~\ref{tab:benchmark-points-exp} as true values.
Our first benchmark is the current global best fit point for normal
mass ordering~\cite{deSalas:2017kay}. The current global fit results
for normal ordering also allow for a local minimum in the upper octant
of $\theta_{23}$ as listed in Tab.~\ref{tab:benchmark-points-exp}.
As our second experimentally motivated benchmark, we took this as the
true value for the DUNE simulation. The results of our simulations for
the global (left panel) and local minimum (right panel) are shown in
Fig.~\ref{fig:sq23-del-GLOBAL-FIT-NO}.

As can be seen in the left panel of
Fig.~\ref{fig:sq23-del-GLOBAL-FIT-NO}, if the true values of
$\theta_{23}$ and $\delta_{CP}$ correspond to the current best fit
value, then after only two years of running time, DUNE will be able to
probe a large part of the parameter space at 3$\sigma$. 
After seven years of running time, the DUNE sensitivity will be much
higher, so that at 3$\sigma$ C.L. it will rule out the possibility of
$\theta_{23}$ being maximal, or lying in the upper octant. 
It will also rule out all possibilities of no CP violation in the
lepton sector. At 5$\sigma$, apart from a very small parameter range,
one finds that both the upper octant of $\theta_{23}$ as well as the
CP conserving case will be ruled out. The maximal atmospheric angle
can be ruled out at even higher confidence level.\\[-.3cm]

On the other hand, for the current local minimum, after just two years
of running DUNE will be able to rule out the lower octant solutions at
3$\sigma$ C.L. except for a small region of parameters, as can be seen
in the right panel of Fig.~\ref{fig:sq23-del-GLOBAL-FIT-NO}.
With seven years of running time, the allowed region in the
$\theta_{23}-\delta_{CP}$ plane shrinks considerably, so that
the left octant appears only at 5$\sigma$ C.L. 
The maximal value of $\theta_{23}$ will be ruled out at higher significance.\\[-.3cm]

\begin{figure}[!t]
 \centering
      \includegraphics[width=0.7\textwidth]{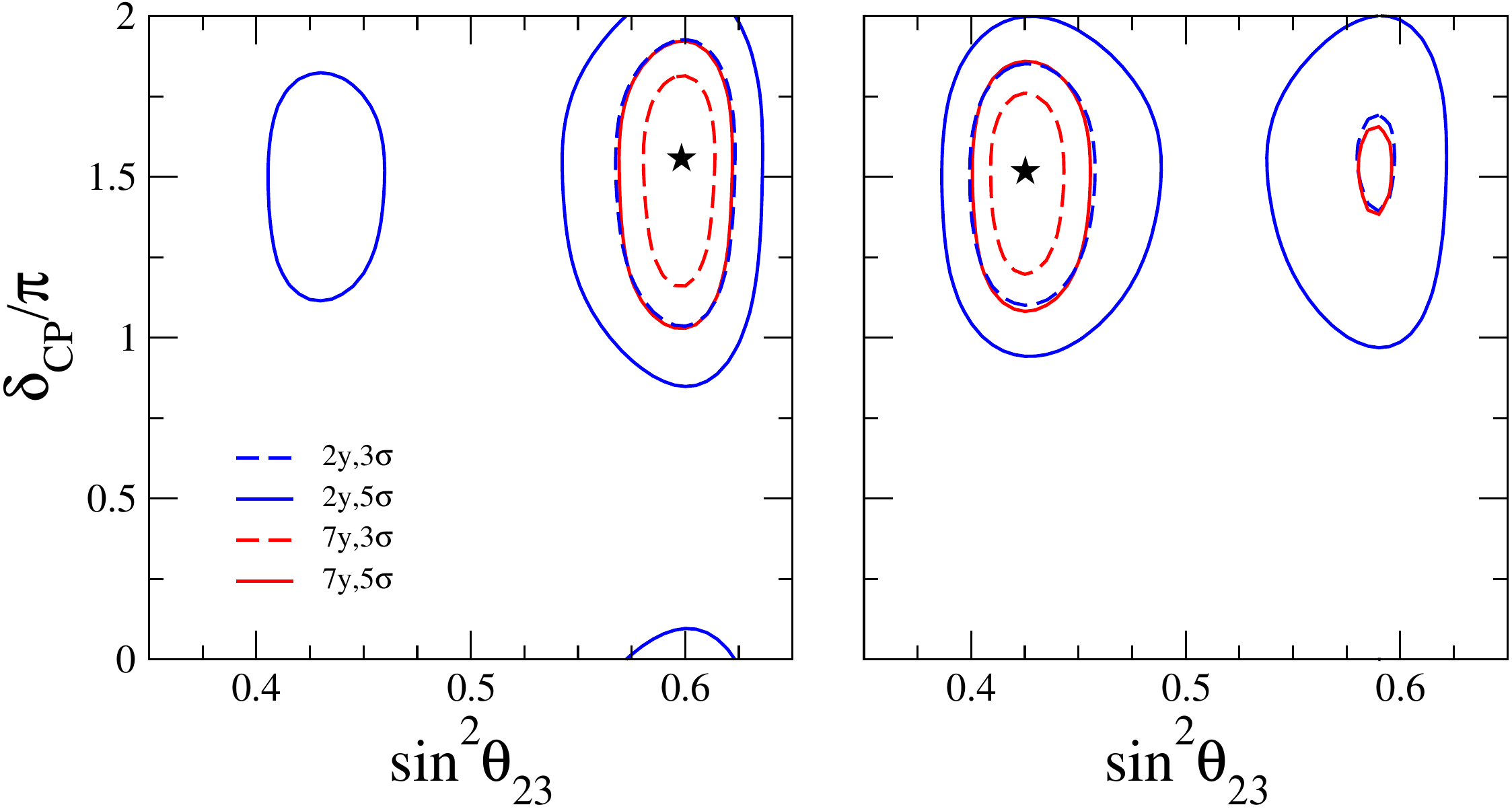}
      \captionsetup{justification=raggedright}
      \caption{DUNE sensitivity projections for inverted mass ordering
        of neutrinos, after 2 years (blue) and 7 years (red) run,
        taking the global best fit point (left panel) and local
        minimum (right panel) of Ref.~\cite{deSalas:2017kay} as
        benchmark points (shown in star).  }
	\label{fig:sq23-del-GLOBAL-FIT-IO}
\end{figure}

Since we currently do not know the mass ordering of
neutrinos~\cite{Gariazzo:2018pei}, we have also examined the case of
inverted ordering (IO). As in the previous case, for inverted ordering
there are also two minima. The current data give a preferred minimum
in the second octant, as well a local minimum in the first octant of
$\theta_{23}$.
The results of our DUNE simulation for these benchmark points are
shown in Fig.~\ref{fig:sq23-del-GLOBAL-FIT-IO}.
The left panel shows our DUNE simulation results for inverted mass
ordering taking the global minimum as true value. After running for
two years, most of the $\theta_{23}-\delta_{CP}$ plane, including
maximal mixing and lower octant of $\theta_{23}$, would appear only
at 5$\sigma$ C.L. 
Moreover, the  CP conserving hypothesis will be disfavored at more
than 5$\sigma$ after seven years running time. 
The right panel shows the results of our simulation for the IO case
when we take the local minimum as true value. Again, DUNE measurements
will considerably shrink the allowed region in the
$\theta_{23}-\delta_{CP}$ plane. 
After two years of running time,  CP conservation and maximal
atmospheric mixing could be excluded beyond the 3$\sigma$ level, while
only a very small region of parameters for the upper octant solution would survive.
After seven years of running time, the upper octant solution would appear
only at 5$\sigma$ C.L., while  maximal atmospheric mixing and CP conservation
would be ruled out beyond 5$\sigma$. 

 In short, should any of the current experimental benchmark points
be the true value of neutrino oscillation parameters, then DUNE would
exclude a very large part of the $\theta_{23}-\delta_{CP}$ plane,
including opposite octant solutions, CP conserving scenarios as well as
maximal mixing. 

After taking a detailed look at the DUNE capabilities for
experimentally motivated benchmark points, we now turn to the
theoretically motivated scenarios. As mentioned before, one of the
frequently occurring predictions in different models of leptonic
mixing is the maximal mixing with CP conservation. 
There are two possible benchmark points corresponding to such a
scenario, namely ($\sin^2 \theta_{23} = 0.5, \delta_{CP} = 0$) and
($\sin^2 \theta_{23} = 0.5, \delta_{CP} = \pi$). 
We took these two points as our first theory motivated benchmark
points for our DUNE simulations. 
\begin{figure}[h!t]
 \centering
      \includegraphics[width=0.7\textwidth]{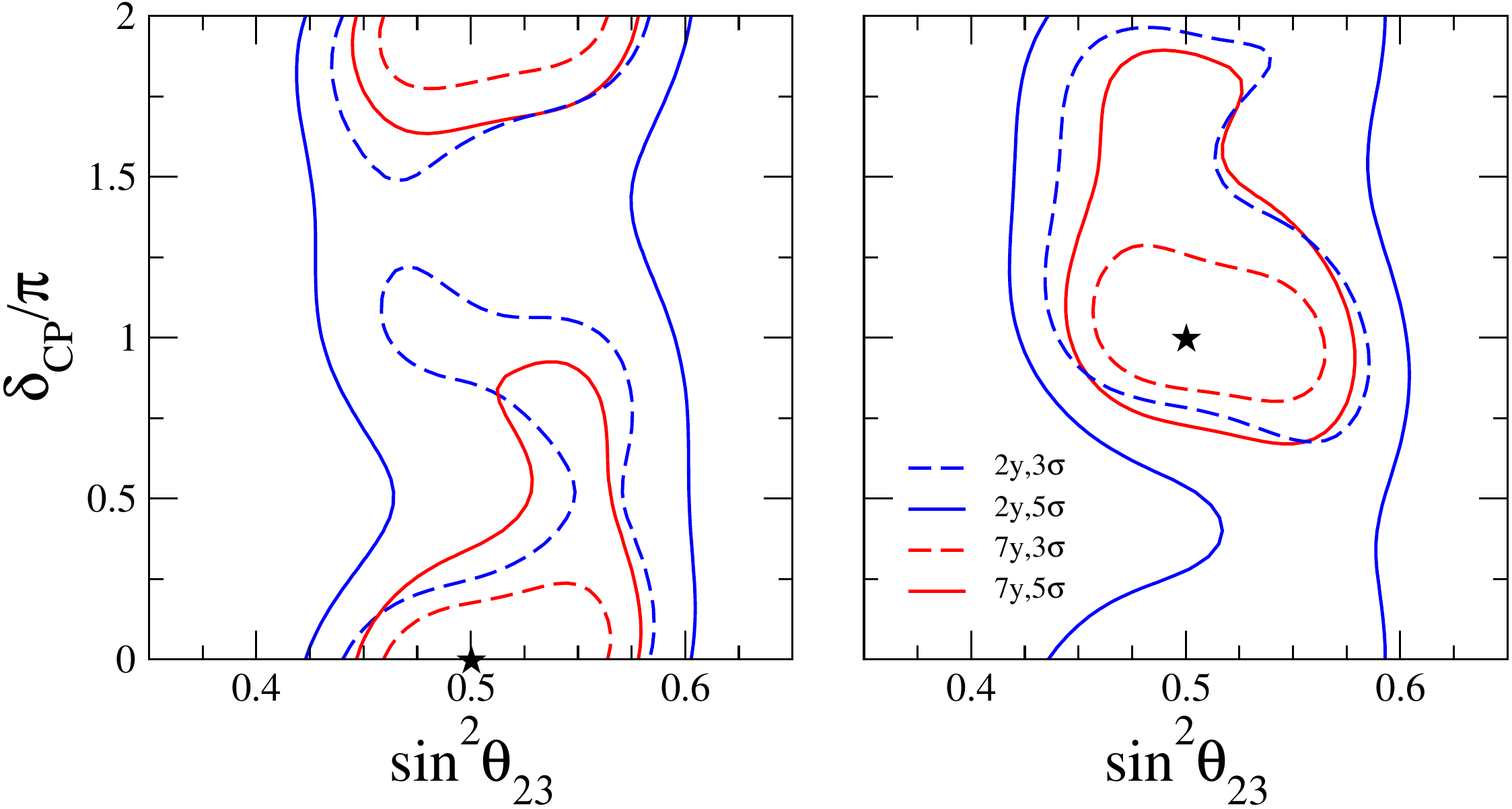}
      \captionsetup{justification=raggedright}
      \caption{DUNE sensitivity projections expected after 2 years
        (blue) and 7 years (red) run, taking maximal mixing
        ($\sin^2 \theta_{23} = 0.5$) and no CP violation
        with $\delta_{CP} = 0$ (left panel) and $\delta_{CP} = \pi$
        (right panel) as benchmark points (star).  }
	\label{fig:sq23-del-TBM}
\end{figure}
In the left panel of Fig.~\ref{fig:sq23-del-TBM} we have taken
($\sin^2 \theta_{23} = 0.5, \delta_{CP} = 0$) as the true value used to generate DUNE data. 
After two years of running time, the allowed region will shrink appreciably,
particularly for $\theta_{23}$, although the maximal CP violation will
still be allowed at 3$\sigma$ C.L.
After seven years, the allowed region will shrink much further, so that
maximal CP violation will be ruled out at 3$\sigma$ C.L, though  a small parameter 
region  will still remain  allowed at 5$\sigma$.
In the right panel of Fig.~\ref{fig:sq23-del-TBM} we have taken the
other CP conserving point,
($\sin^2 \theta_{23} = 0.5, \delta_{CP} = \pi$), as true value.  With two
years of running, DUNE will probe a large parameter region at
3$\sigma$ C.L., although maximal CP violation would be still allowed at that confidence level.
After seven years of running time, however, the possibility of maximal
CP violation would be excluded at 3$\sigma$, and the allowed region
for $\theta_{23}$ will further shrink.
At 5$\sigma$ C.L. only a small region for maximal CP violation
would still survive. 

\begin{figure}[b]
 \centering
      \includegraphics[width=0.7\textwidth]{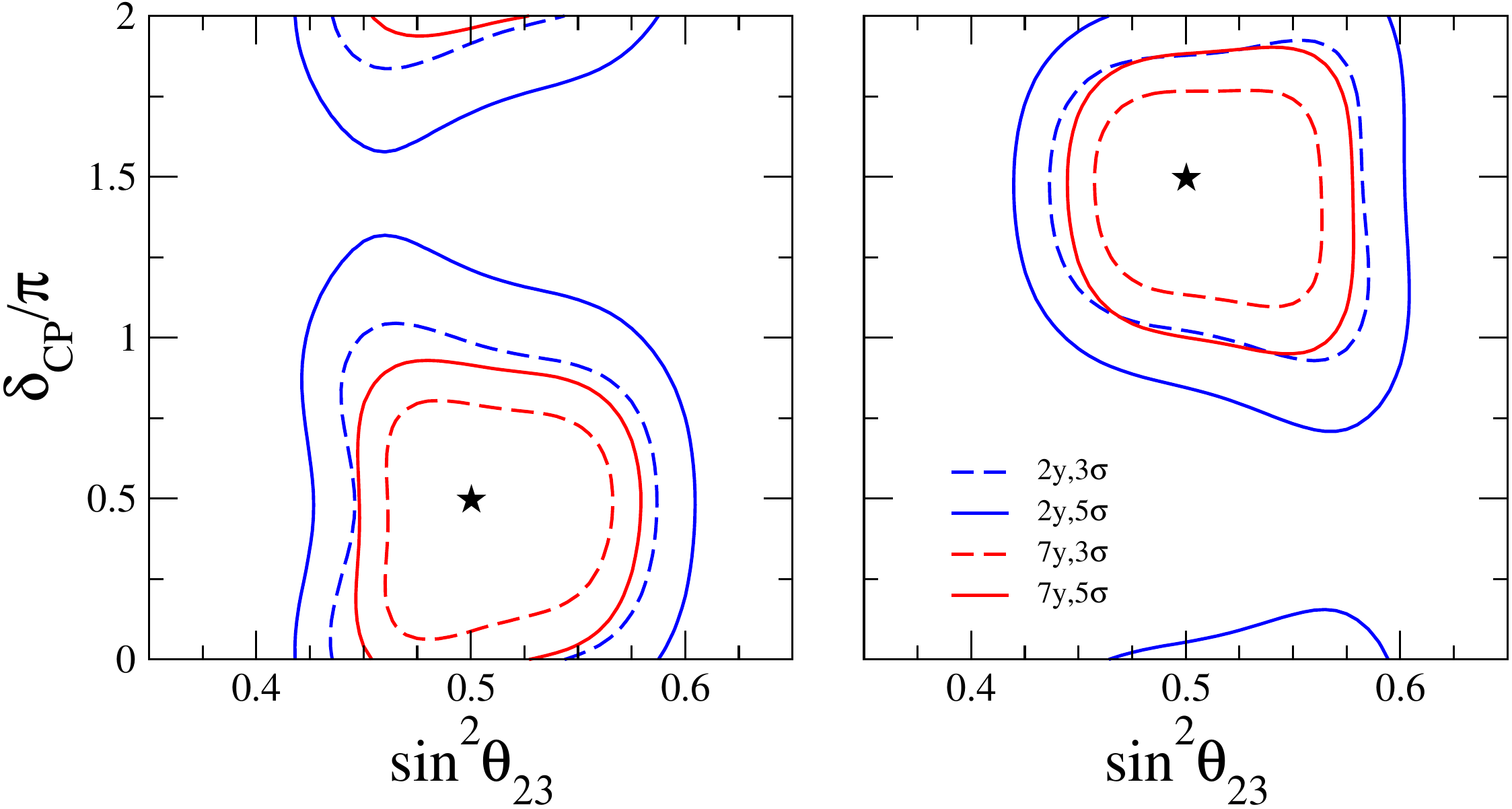}
      \captionsetup{justification=raggedright}
      \caption{DUNE sensitivity projections expected after 2 years (blue) and 7
        years (red) run, taking maximal mixing
        ($\sin^2 \theta_{23} = 0.5$) and maximal CP violation
        i.e. $\delta_{CP} = \pi/2$ (left) and
        $\delta_{CP} = 3\pi/2$ (right panel) as benchmark
        points (star).  }
	\label{fig:sq23-del-COBI}
\end{figure}

Another theoretically well motivated case is the one of maximal
atmospheric mixing and maximal CP violation. As in the previous case,
two possibilities arise here, namely
($\sin^2 \theta_{23} = 0.5, \delta_{CP} = \pi/2$) and
($\sin^2 \theta_{23} = 0.5, \delta_{CP} = 3\pi/2$). As our next
example we have considered these two possibilities as true values, with
the results shown in Fig. \ref{fig:sq23-del-COBI}.
The left panel of Fig.~\ref{fig:sq23-del-COBI} considers the former point as true value of the oscillation 
parameters, while the latter one has been considered in the right panel. In both cases, after two years of running time, the allowed
region will shrink considerably. 
However, the possibility of  CP conservation would remain allowed at 3$\sigma$. The
situation improves significantly after analyzing seven years of DUNE running
time.
In that case, the CP conservation hypothesis would be completely
excluded for both benchmark points at 3$\sigma$ C.L. At 5$\sigma$ only a small
parameter region for CP conservation would survive in both cases.

\begin{figure}[!t]
 \centering
      \includegraphics[width=0.35\textwidth]{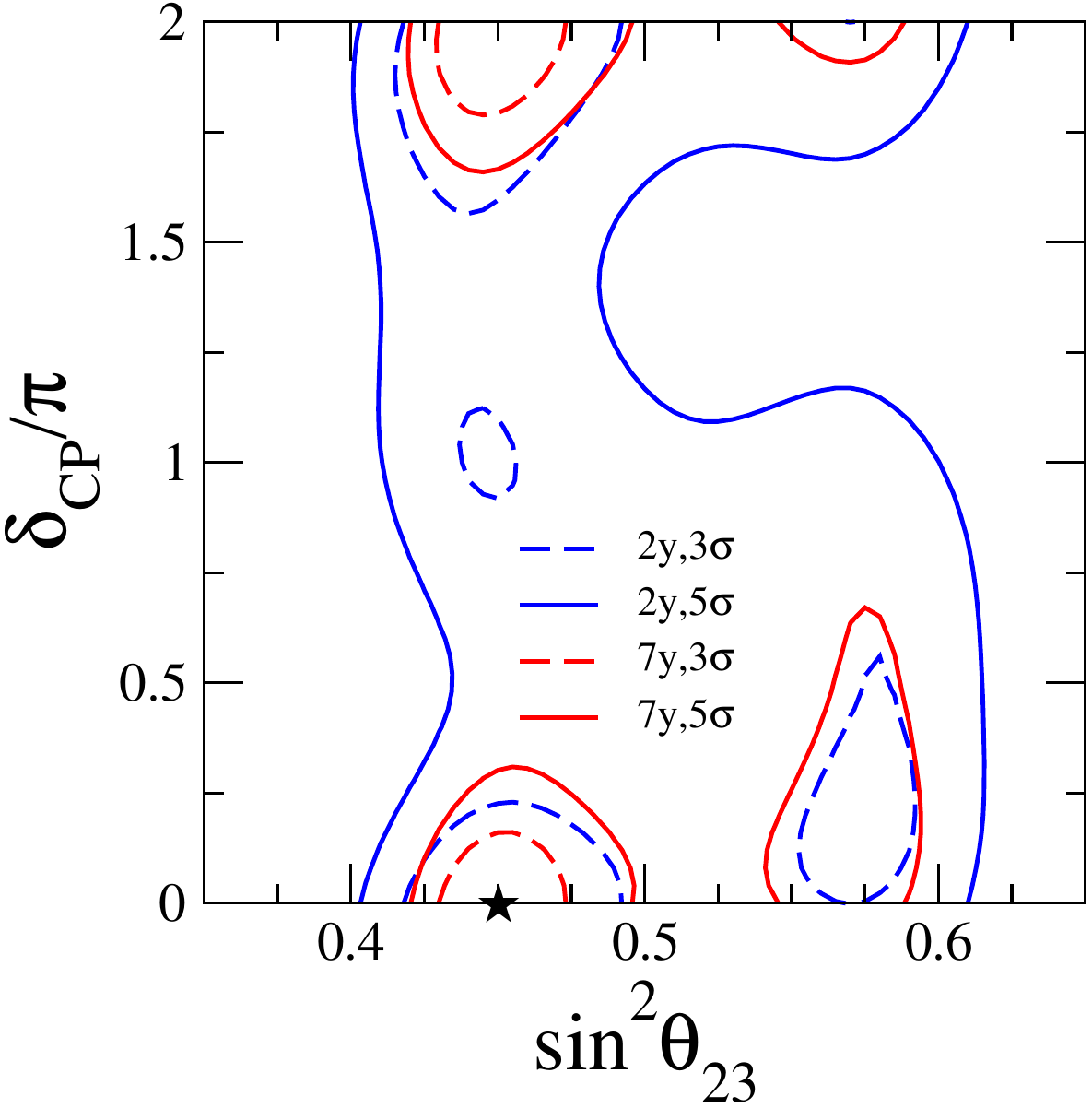}
      \captionsetup{justification=raggedright}
      \caption{DUNE sensitivity projections expected after 2 years
        (blue) and 7 years (red) run, for the CP conserving bi-large
        mixing scenario ($\sin^2 \theta_{23} =
        0.45$,~$\delta_{CP} = 0$) as our benchmark point (star).  }
	\label{fig:sq23-del-BILARGE}
\end{figure}

As our final theory motivated benchmark point we examined the bi-large
mixing scenario~\cite{Boucenna:2012xb,Ding:2012wh,Roy:2014nua} as the
true value for our DUNE simulation.
As explained before, this benchmark corresponds to
$\sin^2 \theta_{23} = 0.45$. 
 While the bi-large mixing ansatz in its
simplest form does not predict any particular value of $\delta_{CP}$,
here we have taken bi-large mixing without  CP
violation, $\delta_{CP} = 0$, as our reference choice. The result of our
simulation in this case is shown in Fig.~\ref{fig:sq23-del-BILARGE}.
Also here, as can be seen in the figure, after two
years of DUNE data taking, the allowed parameter region will shrink
considerably. 
However, the possibility of upper octant values of $\theta_{23}$ will
still be allowed in some part of the parameter space at 3$\sigma$.
As expected, after seven years of DUNE data taking, there will substantial
improvement in the sensitivity to both parameters. 
Indeed, the upper octant solution for $\theta_{23}$ and maximal CP violating
scenarios will be completely excluded  at
3$\sigma$. At 5$\sigma$, maximal CP violation will only be allowed in a
small region of upper octant values of $\theta_{23}$.

\subsection{Line Benchmarks}
\label{sec:line-benchmarks}

So far we have only considered benchmark points. In this section we
consider more model-independent scenarios associated to line-like
cases as possible true values in our simulations.  The three line-like
benchmark cases under study are: 
\begin{enumerate}
 \item Maximal atmospheric mixing ($\sin^2 \theta_{23} = 0.5$) for all possible values of $\delta_{CP}$.
 \item CP conservation ($\delta_{CP} = 0 \,, \pi$) with arbitrary values of  $\theta_{23}$.
 \item Maximal CP violation ($\delta_{CP} = \pi/2 \,, 3\pi/2$) with arbitrary values of  $\theta_{23}$.
\end{enumerate}

\begin{figure}[!t]
 \centering
      \includegraphics[width=0.35\textwidth]{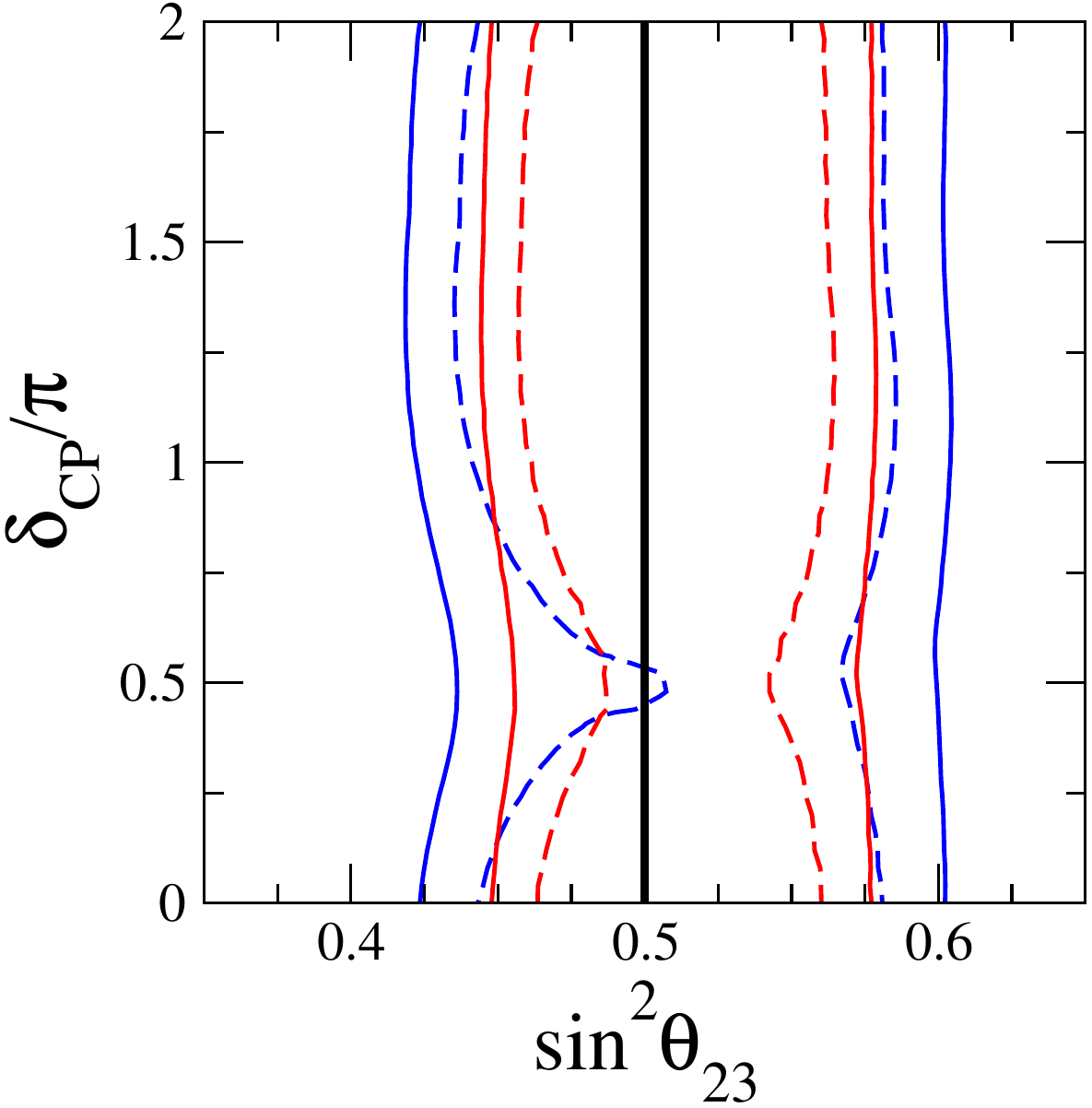}
      \captionsetup{justification=raggedright}
      \caption{DUNE sensitivity projections after 2 years
        (blue) and 7 years (red) run, taking the maximal mixing
        ($\sin^2 \theta_{23} = 0.5$) scenario (arbitrary
        $\delta_{CP}$) as benchmark line (black).  }
	\label{fig:sq23-del-LINE-th23}
\end{figure}

As discussed in Sec.~\ref{sec:benchmark}, our first benchmark line
occurs frequently in flavor models of leptonic mixing.  The result of
the simulation for this benchmark line is shown in
Fig.~\ref{fig:sq23-del-LINE-th23}.
As can be seen, after two years of running time, DUNE will
considerably narrow down the currently allowed parameter range at both
3$\sigma$ and 5$\sigma$ C.L. 
After seven years of running time, the
allowed region of parameter space for $\theta_{23}$ will shrink even
further. Since in our simulations we have taken all possible values of
$\delta_{CP}$ along the $\sin^2 \theta_{23} = 0.5$ line as true
values, the information about the DUNE probing capabilities with
respect to $\delta_{CP}$ is naturally lost in this simulation. 
Notice that the small kink in the 3$\sigma$ curve of our simulation
for two year run around $\delta_{CP} = \pi/2$ is understood
and reflects the fact that our simulations take into account the
current experimental knowledge on these parameters which 
disfavors $\delta_{CP} \approx \pi/2$. 
As expected, after seven year of running time the kink is much less
pronounced because at this point the corresponding curves for these
parameters are totally driven by DUNE, hence the effect of the
current global fit is washed out. \\[-.2cm]

\begin{figure}[!t]
 \centering
      \includegraphics[width= 0.7\textwidth]{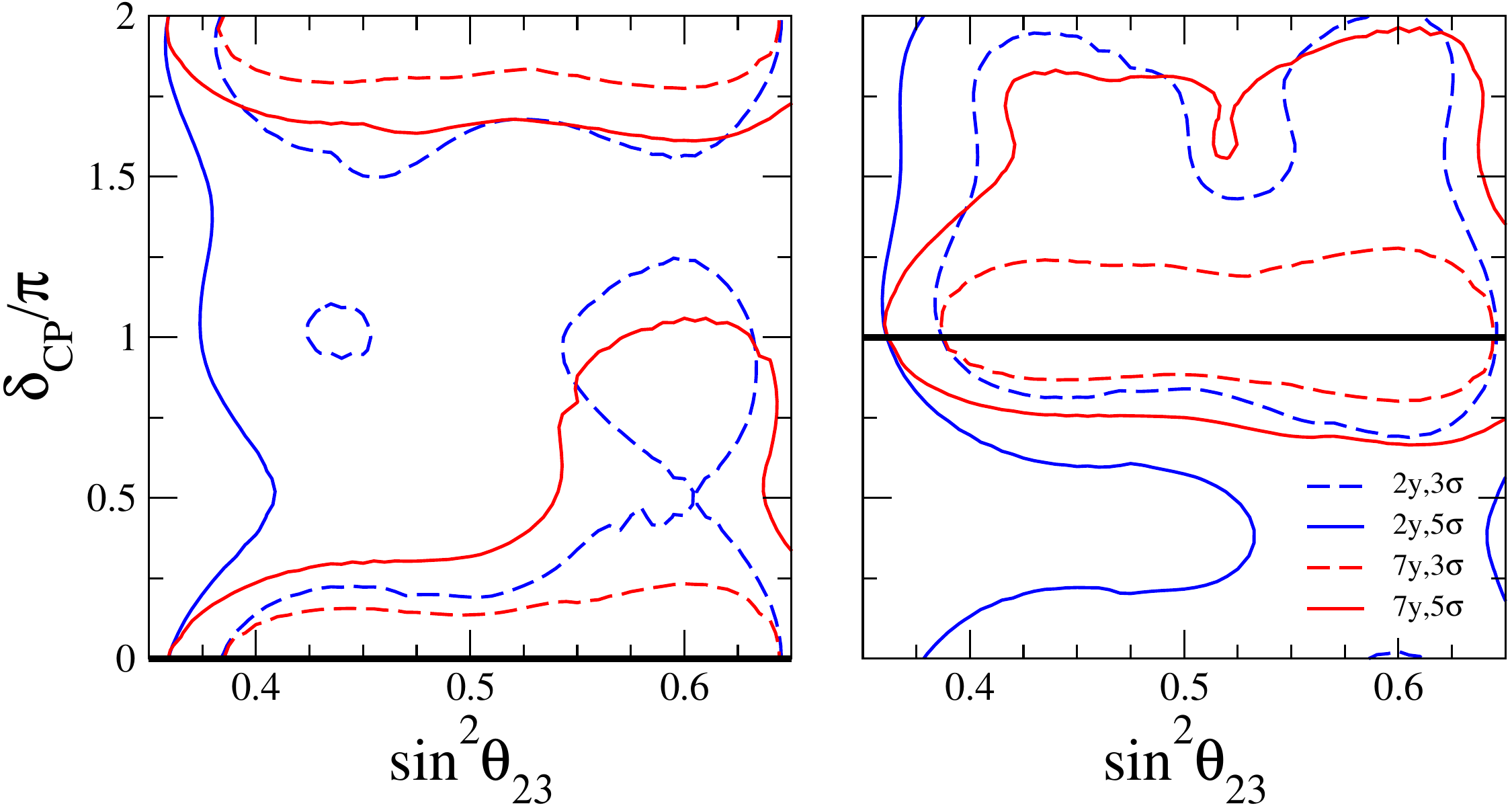}
      \captionsetup{justification=raggedright}
      \caption{DUNE sensitivity projections expected after 2 years
        (blue) and 7 years (red) run, taking no CP violation cases
        $\delta_{CP} = 0$ (left panel) and
        $\delta_{CP} = \pi $ (right panel) for arbitrary
        $\sin^2 \theta_{23}$ as the benchmark lines (black).  }
	\label{fig:sq23-del-LINE-NOCP}
\end{figure}

The next benchmark line we have examined is that corresponding to the
CP conserving hypothesis, i.e., $\delta_{CP} = 0$ or $\delta_{CP} = \pi$.
We took these two values as benchmark lines for the rather conservative
range of $\sin^2\theta_{23} \in [0.35,0.65]$, as
mentioned in Tab.~\ref{tab:benchmark-lines}. 
The results for this case are shown in Fig.~\ref{fig:sq23-del-LINE-NOCP}.
As can be seen in the left panel,
corresponding to $\delta_{CP} = 0$, after just two years of data
collection, DUNE can severely constrain the allowed range of
$\delta_{CP}$ at 3$\sigma$ C.L.
In fact, the possibility of maximal CP violation is almost ruled out at
3$\sigma$ C.L.
After seven years of running time, the allowed region will shrink much
more and DUNE would be able to  exclude completely the possibility of maximal CP
violation at 3$\sigma$ C.L. 
Even at 5$\sigma$ C.L., the allowed region after seven years of run would be
significantly reduced and, apart from a small region, the possibility
of maximal CP violation will be essentially excluded. 
The right panel of Fig.~\ref{fig:sq23-del-LINE-NOCP} shows the result
of our simulation for $\delta_{CP} = \pi$. 
Again, after two years of DUNE running time, at 3$\sigma$ C.L. the
allowed region $\delta_{CP}$ will shrink considerably. However, unlike
in the previous case, here the possibility of maximal CP violation
will still be allowed for most of the $\theta_{23}$ range. 
The root for this loss of discriminating power with respect to the previous
case is the fact that the current global fit data prefers CP
violation close to $\delta_{CP} \approx 3\pi/2$. 
Thus, in this case, our two year run DUNE simulations (which include
the current global fit results) would not be able to rule out maximal CP violation.
In contrast, however, after seven years of running time of DUNE, the
simulation is mainly driven by the DUNE data sample that
 will be able to, not only further shrink the parameter space, but also  rule out maximal CP
violation at 3$\sigma$ C.L. 
Before moving on to next case, we would like to remark that, since we have
taken the whole range of $\theta_{23}$ as true value for our benchmark
lines, naturally the information about DUNE reach to $\theta_{23}$ is
lost in this simulation. However, note that the extreme values of
$\theta_{23}$ in both panels of Fig. \ref{fig:sq23-del-LINE-NOCP} are
indeed getting excluded in both two and seven year runs as these edge
values are disfavored by current global fits at a very high significance. \\[-.2cm] 

\begin{figure}[!t]
 \centering
      \includegraphics[width=0.7\textwidth]{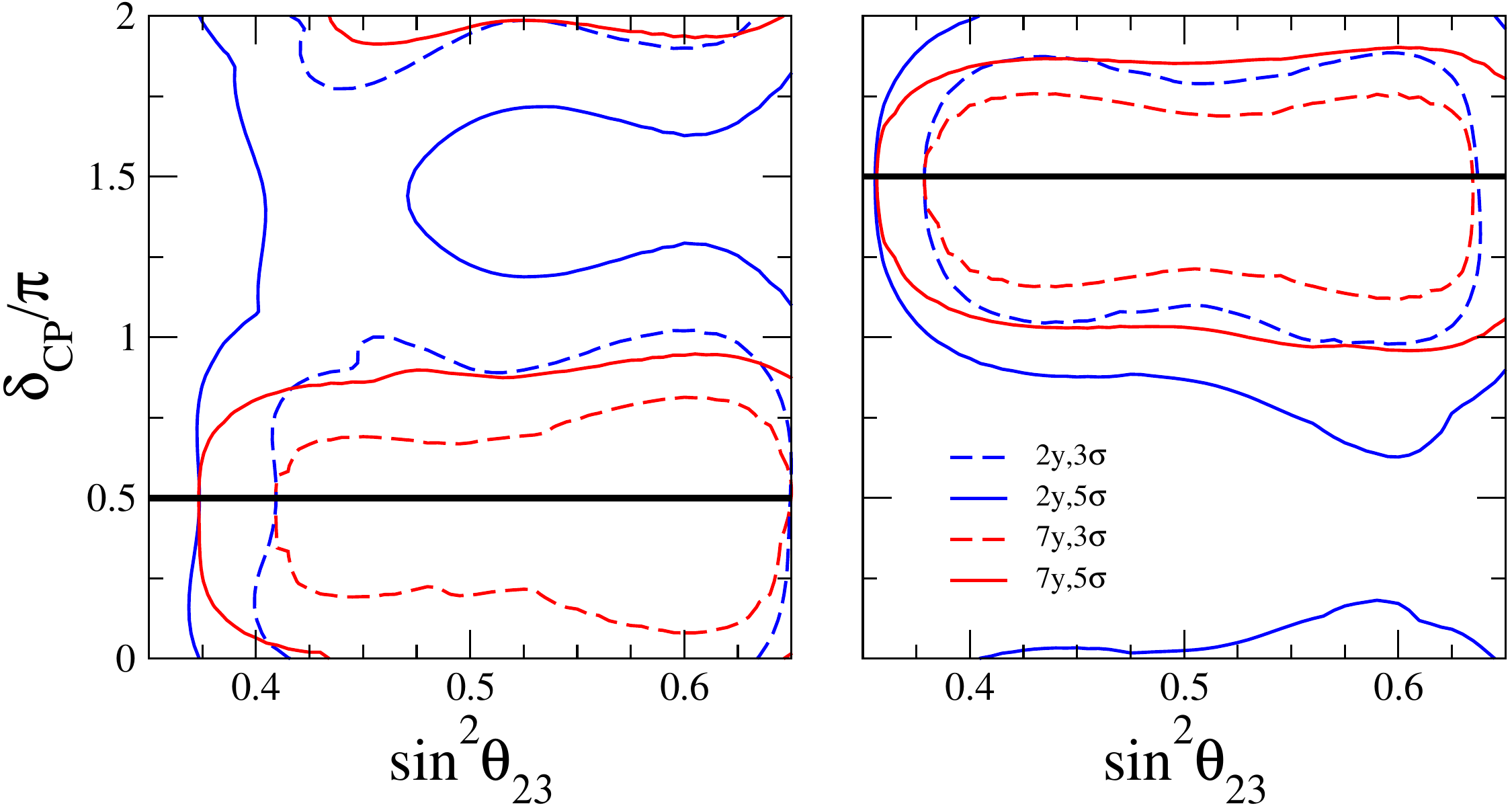}
      \captionsetup{justification=raggedright}
      \caption{ DUNE sensitivity projections expected after 2 years
        (blue) and 7 years (red) run, taking maximal CP violation
        cases $\delta_{CP} = \pi/2$ (left) and
        $\delta_{CP} = 3\pi/2$ (right panel) for
        arbitrary $\sin^2 \theta_{23}$ as the benchmark lines
        (black).  }
	\label{fig:sq23-del-LINE-MAXCP}
\end{figure}

As our final benchmark line, we have looked at the possibility of
maximal CP violation. There are several theoretically motivated models
that predict such a case. Also, the current experimental data
indicates nearly maximal CP violation with
$\delta_{CP} \approx 3\pi/2$. 
Again there are two possibilities for maximal CP violation namely
$\delta_{CP} = \pi/2$ or $\delta_{CP} = 3\pi/2$. We have
taken  these two values as our benchmark lines in the
simulation for all allowed values of $\theta_{23}$ in the range
$\sin^2 \theta_{23} = [0.35,0.65]$. The result is shown in
Fig.~\ref{fig:sq23-del-LINE-MAXCP}. 
The left panel of this figure corresponds to
the case of benchmark line $\delta_{CP} = \pi/2$.
From the plot one can see that, after two years of running time of DUNE, the
allowed values for $\delta_{CP}$ will shrink appreciably. 
 However, the possibility of no CP violation will still be allowed
for most  values of $\theta_{23}$  at the 3$\sigma$ level.
After seven years of running time the situation will  drastically improve, and the  CP conserving hypothesis
could be ruled out at 3$\sigma$ C.L. 
The right panel of Fig.~\ref{fig:sq23-del-LINE-MAXCP} shows the
results for the benchmark line with $\delta_{CP} = 3\pi/2$. 
In this case, after two years of running time, DUNE will be able to rule
out the  CP conserving scenario at 3$\sigma$ C.L. in the whole
parameter range. 
Again, our results for two year simulations can be understood as
arising from the fact that our simulations take into account the
current experimental information on $\delta_{CP}$, as explained before. 
After seven years of running time, DUNE could considerably restrict the
allowed range for $\delta_{CP}$, ruling out the CP conserving scenario 
at 5$\sigma$.
As before, note that, since we have taken the whole range of
$\sin^2 \theta_{23} \in [0.35,0.65]$ as true value, the DUNE
sensitivity on $\theta_{23}$ is not apparent in the simulations. 
However, as shown in both panels, the extreme end values of $\theta_{23}$, currently
strongly disfavored by  oscillation data, can be totally excluded at the 5$\sigma$ level.

 A comment is in order concerning the benchmark line results, namely,
 the fact that in nature the true value of a given neutrino
 oscillation parameter will always correspond to a single point in
 parameter space.
Nonetheless, the line-like simulations do carry useful information and
can be used as a guide to narrow down the actual allowed range of these
parameters. 
However, since in these simulations one takes all possible values
of a given parameter lying on a line as true values, naturally the
results of these simulations will not be as constraining as the
results obtained in the previous section.


\section{Summary and discussion}
\label{sec:summary-discussion}


Using the design specifications of the DUNE experiment and taking into
account the current status of neutrino oscillation parameters, as
summarized in global oscillation fits, we have determined DUNE's
potential to probe the pattern of neutrino mixing and CP violation
after two and seven years of running.
We have taken various input benchmark values as our true values.
These include not only the current preferred values of $\theta_{23}$
and $\delta_{CP}$, as given in global oscillation fits, but also
several theory-motivated choices as to what the true values of
$\theta_{23}$ and $\delta_{CP}$ could be.
We have examined quantitatively DUNE's capability to probe deviations
from maximality in the atmospheric angle $\theta_{23}$ and its octant,
as well as to probe for the CP violation hypothesis itself, in a
model-independent way.
We have found that, after seven years of running, DUNE will be able
to test quantitatively the predictions of various models of neutrino
mixing.
Our results are summarized in Figs.~\ref{fig:sq23-del-GLOBAL-FIT-NO}-\ref{fig:sq23-del-LINE-MAXCP}.
They can be used so as to establish DUNE's capability to
discriminate between the various mixing scenarios analyzed above.
Table~\ref{tab:summary}, gives an ``executive summary'' of DUNE's
sensitivity to our various benchmark choices described in
Sec.~\ref{sec:benchmark}. 
\begin{table}[t!]\centering
  \catcode`?=\active \def?{\hphantom{0}}
   \begin{tabular}{|p{1.7cm}|p{1.7cm}|p{1.7cm}|p{1.7cm}|p{1.7cm}|p{1.7cm}|p{1.7cm}|p{1.7cm}|p{1.7cm}|}
    \hline 
\backslashbox{True}{Test} & \makecell{GM\\(NO)} & \makecell{LM\\(NO)} & \makecell{$\theta_{23} = 45\degree$\\$\delta_{CP} = 0$} & \makecell{$\theta_{23} = 45\degree$\\$\delta_{CP} = \pi$} & \makecell{$\theta_{23} = 45\degree$\\$\delta_{CP} = \frac{\pi}{2}$} & \makecell{$\theta_{23} = 45\degree$\\$\delta_{CP} = \frac{3\pi}{2}$} & \makecell{Bi-large}  \\
\hline
\makecell{GM\\(NO)}  &  \makecell{-}  & \makecell{ \cmark(\xmark)}    & \makecell{\cmark(\cmark)}  & \makecell{\cmark(\cmark)}  & \makecell{\cmark(\cmark)}  & \makecell{\cmark(\cmark)}  & \makecell{\cmark(\cmark)}  \\ 
\hline
\makecell{LM\\(NO)}  & \makecell{\cmark(\xmark)}    &  \makecell{-}  & \makecell{\cmark(\cmark)}  & \makecell{\cmark(\cmark)}  & \makecell{\cmark(\cmark)}  & \makecell{\cmark(\cmark)}  &  \makecell{\cmark(\cmark)} \\ 
\hline
\makecell{$\theta_{23} = 45\degree$\\$\delta_{CP} = 0$}         &  \makecell{\cmark(\cmark)}  &  \makecell{\cmark(\cmark)}     & \makecell{-} & \makecell{\cmark(\cmark)}  &  \makecell{\cmark(\cmark)} & \makecell{\cmark(\cmark)}  & \makecell{\cmark(\xmark)}  \\ 
\hline
\makecell{$\theta_{23} = 45\degree$\\$\delta_{CP} = \pi$}       &  \makecell{\cmark(\cmark)}  &  \makecell{\cmark(\cmark)}     & \makecell{\cmark(\cmark)}  & \makecell{-} & \makecell{\cmark(\cmark)}  &  \makecell{\cmark(\xmark)} & \makecell{\cmark(\cmark)}  \\ 
\hline
\makecell{$\theta_{23} = 45\degree$\\$\delta_{CP} = \frac{\pi}{2}$}   & \makecell{\cmark(\cmark)}   &  \makecell{\cmark(\cmark)}     & \makecell{\cmark(\xmark)}  &  \makecell{\cmark(\cmark)} & \makecell{-} & \makecell{\cmark(\cmark)}  &  \makecell{\cmark(\cmark)} \\ 
\hline
\makecell{$\theta_{23} = 45\degree$\\$\delta_{CP} = \frac{3\pi}{2}$}   &  \makecell{\cmark(\cmark)}  & \makecell{\cmark(\cmark)}      & \makecell{\cmark(\cmark)}  &  \makecell{\cmark(\xmark)} &  \makecell{\cmark(\cmark)} & \makecell{-} &  \makecell{\cmark(\cmark)} \\
\hline
\makecell{Bi-large \\ }                                                   &  \makecell{\cmark(\cmark)}  & \makecell{\cmark(\cmark)}      & \makecell{\cmark(\cmark)}  & \makecell{\cmark(\cmark)}  & \makecell{\cmark(\cmark)}  & \makecell{\cmark(\cmark)}  & \makecell{-}  \\   
    \hline
     \end{tabular}
          \captionsetup{justification=raggedright}
          \caption{The discriminating power of DUNE after running for seven years. 
            Various mixing hypothesis, taken as true, are listed in the rows, while the columns 
            indicate the mixing hypotheses that can be tested against the assumed true scenario. The
            convention for ticks and crosses is given in the text. }
     \label{tab:summary} 
\end{table}
Each row in Tab.~\ref{tab:summary} corresponds to an assumed pair of
parameters ($\theta_{23}$, $\delta_{CP}$), taken as true values, while
the columns indicate the different mixing hypotheses that can be
tested from the simulated DUNE data.
The table corresponds to the results obtained for the case of normal
mass ordering, from the simulation of seven years of future DUNE
neutrino oscillation data, with a confidence level of 3$\sigma$ and
5$\sigma$ C.L. (in parentheses).  Ticks and crosses mean that,
assuming the true oscillation parameters as given in each row, the
particular benchmark shown in every column can be ruled out at the
given confidence, or not, respectively.
For example, if we assume as true value for the oscillation parameters
$\theta_{23}$ and $\delta_{CP}$ the ones given by the global minimum
(GM) of the current oscillation fit (see
Tab.~\ref{tab:benchmark-points-exp}), after seven years or run, DUNE
will be able to exclude all the other analyzed mixing scenarios at
3$\sigma$ and 5$\sigma$ C.L., except for the local minimum (LM)
benchmark point, that will be still compatible with data at the
5$\sigma$ level.
Moreover, one sees that, with seven years of data, DUNE will have
enough sensitivity to discriminate among all the benchmark points
analyzed at 3$\sigma$ C.L., with only a few ambiguities remaining at
the 5$\sigma$ level.
In summary, one can conclude that DUNE will make a substantial step
towards the precise determination of ($\theta_{23}$, $\delta_{CP}$),
bringing to quantitative test the predictions of various theories of
neutrino mixing.

\begin{acknowledgments}

  Work supported by the Spanish grants FPA2017-85216-P and
  SEV-2014-0398 (MINECO), and PROMETEOII/2014/084 and GV2016-142 grants from Generalitat
  Valenciana.
MT is also supported by a Ram\'{o}n y Cajal contract (MINECO).
CT is supported by the FPI fellowship BES-2015-073593 (MINECO).

\end{acknowledgments}



\end{document}